\documentclass[article, shortnames]{jss}

\usepackage{thumbpdf,lmodern}

\usepackage{framed}
\usepackage{amssymb}
\usepackage{amsmath}
\usepackage{bbm}
\usepackage[ruled,linesnumbered]{algorithm2e}

\author{Oluwasegun Ojo\\IMDEA Networks Institute\\Universidad Carlos III de Madrid
   \And Rosa E. Lillo  \\  Department of Statistics \\ Universidad Carlos III de Madrid \\ uc3m-Santander Big Data Institute
   \AND Antonio Fern\'andez Anta \\ IMDEA Networks Institute }
\Plainauthor{Oluwasegun Ojo,  Rosa E. Lillo, Antonio Fernandez-Anta}

\title{Outlier Detection for Functional Data with \proglang{R} Package \pkg{fdaoutlier}}
\Plaintitle{Outlier Detection for Functional Data with R Package fdaoutlier}

\Abstract{
Outlier detection is one of the standard exploratory analysis tasks in functional data analysis. We present the \proglang{R} package \pkg{fdaoutlier} which contains implementations of some of the latest techniques for detecting functional outliers. The package makes it easy to detect different types of outliers (magnitude, shape, and amplitude) in functional data, and some of the implemented methods can be applied to both univariate and multivariate functional data. We illustrate the main functionality of the \proglang{R} package with common functional datasets in the literature.
}

\Keywords{functional data analysis, outlier detection, fdaoutlier, \proglang{R}}
\Plainkeywords{functional data analysis, outlier detection, fdaoutlier, R}

\Address{
  Oluwasegun Ojo\\
  Global Computing Group\\
  IMDEA Networks Institute\\
  \emph{and}\\
  Universidad Carlos III de Madrid\\
  E-mail: \email{oluwasegun.ojo@imdea.org}\\
  URL: \url{https://statimatics.com/about}
}

\begin{document}



\section[Introduction]{Introduction} \label{sec:intro}

Outlier detection is a common task when carrying out exploratory data analysis. Identifying possible outliers is essential during the exploratory analysis process, because outliers can significantly bias any statistical analysis. The process of dealing with identified outliers may also provide new insights into the data generating process's nature. In functional data analysis (FDA), observations are treated as functions observed on a domain. These functional observations can exhibit various outlyingness properties as pointed out by \cite{hubert2015multivariate}. For instance, an observation can be shifted from the mass of the data. Such outliers are referred to as magnitude outliers in the FDA literature. On the other hand, an observation can be a shape outlier because it differs in shape from the mass of the data (even if it lies completely inside the mass of the data). For periodic functional observations, an observation may be outlying because it has an amplitude different from the mass of the data. Finally, any of the aforementioned outlyingness properties can be exhibited by a functional observation in a subset of the domain or all through the domain. Consequently, identifying outliers in FDA is challenging as there are many possible ways a functional observation can exhibit outlyingness.

Much work has been done regarding identifying outliers in the FDA context, with their corresponding software implementations made available in \proglang{R} \citep{Rcore}. A number of these methods have been obvious applications of a notion of functional depths, which induces a centre outward ordering on a sample of curves. For instance, the functional boxplot \citep{sun2011functional} uses the (modified) band depth to define a 50\% central region for the sample of curves with outliers identified as curves lying outside 1.5 times the central region in any part of the domain. In \proglang{R}, the functional boxplot is available in the \pkg{fda} package \citep{fdapackage} with options to use the fast exact (modified) band depth defined by bands of two functions, proposed in \cite{sun2012exact}.


The \pkg{fda.usc} package \citep{fda.usc} in \proglang{R} implements three functional outlier detection methods. The first method, proposed in \cite{febrero2007functional}, uses a likelihood ratio statistics to detect outlying curves (with cutoff determined by a bootstrap procedure). The two other methods identify outliers by comparing the depth values of the functions to a cutoff also obtained by a bootstrap procedure, based on either trimming of suspicious curves or placing more weights on the deeper curves \citep{febrero2008}. These three methods are also implemented in the \pkg{rainbow} package \citep{hanlinshang2011rj}, as well as the functional bagplot and the functional highest density region plot \citep{hyndman2010bagplot}. The \pkg{rainbow} package also contains the integrated square forecast errors method for detecting functional outliers proposed in \cite{hyndmanullah2018} \citep[see also][]{hyndman2010bagplot}.

\cite{nagy07depth} proposed the $j^{th}$ order integrated and infimal depths for identifying shape outliers, with implementations available in the \pkg{ddalpha} package \citep{pokotylo2019ddalpha}. \cite{rousseeuw2018measure} in their work proposed a directional outlyingness (DO) measure, its functional extension (fDO), and the variability of directional outlyingness (vDO). Then, they used the functional outlier map, a scatter plot of the fDO versus vDO, to identify outliers with cutoffs determined by the standardized logarithm of the combined functional outlyingness(LCFO) measure. The functional outlier map can also be used with the adjusted outlyingness (AO) measure proposed in \cite{brys2005robustification} \citep[see also][and \citeauthor{hubert2015multivariate} \citeyear{hubert2015multivariate}]{hubert2008outlier}, rather than the DO measure. These methods are available in the \pkg{mrfDepth} package \citep{mrfdepth}. Finally, the \pkg{roahd} package \citep{ieva2019roahd} contains an implementation of the outliergram method proposed in \cite{arribas2014shape}, as well as its multivariate generalisation proposed in \cite{ieva2020component}.

More recently proposed outlier detection methods include: the directional outlyingness for multivariate functional data proposed in \cite{dai2019directional} and further elaborated into the magnitude-shape plot (MS-Plot) in \cite{dai2018multivariate}; the total variate depth (TVD) and modified shape similarity index (MSS) proposed in \cite{huang2019decomposition}; and the CRO-FADALARA method, based on archetypoids proposed in \cite{vinue2020robust}, and available in the \pkg{adamethods} package \citep{ada}. \cite{dai2020sequential} also proposed detecting and classifying outliers using some sequence of transformations, e.g., shifting a curve to its centre and normalising it using the $L_2$ norm.

The objective of this paper is to describe the \pkg{fdaoutlier} package which aims to extend the available facility for outlier detection (in the FDA context) for \proglang{R}, with implementations of some of the latest outlier detection methods. The \pkg{fdaoutlier} package's main contributions are:

\begin{itemize}
    \item[-] Implementations of the directional outlyingness and MS-Plot outlier detection methods proposed in \cite{dai2019directional} and  \cite{dai2018multivariate}.

    \item[-] An implementation of the TVD and MSS proposed in \cite{huang2019decomposition}. The \pkg{fdaoutlier} implementation of TVD/MSS is written in \proglang{C++} using \proglang{R}'s \code{.Call} interface which leads to significant computational efficiency as TVD and MSS are computationally intensive.

    \item[-] An implementation of the sequential tranformation method described in \cite{dai2020sequential}.
    \item[-] An implementation of the massive unsupervised outlier detection (MUOD) method proposed in \cite{azcorra2018unsupervised}.
    \item[-] Various depth and ordering methods, including extremal depth, one and two-sided extreme rank length depth, directional quantile, among others, useful for ordering functional observations (e.g., in functional boxplots).

\end{itemize}

In the next section, we describe the theoretical background  of  the implemented outlier detection methods and demonstrate their implementations in \pkg{fdaoutlier} using simulated data. In Section 3, we apply \pkg{fdaoutlier} on two common datasets in the FDA outlier detection literature, replicating some of the analyses done in the literature. We then conclude in Section 4 with some remarks and a future outlook of \pkg{fdaoutlier}.



\section{Outlier detection methods} \label{sec:methods}
\begin{figure}[t!]
\centering
\includegraphics{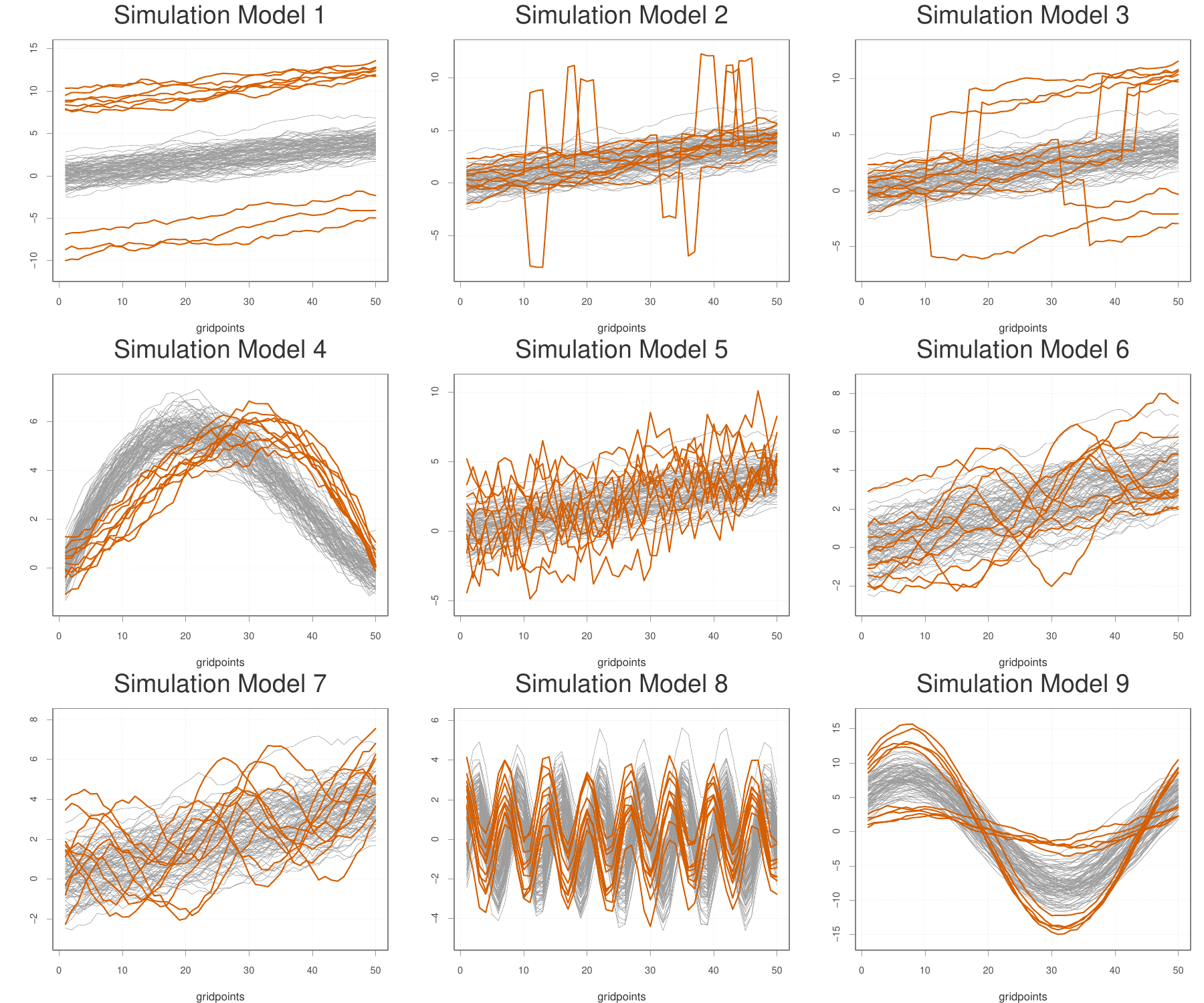}
\caption{\label{fig:simmodels} Simulation models: Plot of sample of data generated by each simulated model in \pkg{fdaoutlier}. Curves in orange are outliers.}
\end{figure}

We provide a brief primer of the implemented methods in the \pkg{fdaoutlier} package, and then describe their implementations. For illustrating the methods, we use the convenience functions \code{simulation_model1()} - \code{simulation_model9()} implemented in \pkg{fdaoutlier} to generate data with different types of outliers. These functions are useful for the rapid development and testing of new outlier detection methods and were curated from the functional outlier detection literature. Figure~\ref{fig:simmodels} shows plots of sample data generated by these nine models produced by calling \code{simulation_model*(plot = TRUE)}.

\subsection{Directional outlyingness and MS-Plot}\label{subsec:2.1}
The directional outlyingess for multivariate functional data proposed in \cite{dai2019directional} provides a
way to measure not only the point-wise outlyingness of a functional observations but also the direction of outlyingness of that observation with respect to (w.r.t.) the rest of the data. Formally, let $\textbf{Y}:I \rightarrow \mathbb{R}^p$ be a stochastic process in the space of real continuous functions $C(I, \mathbb{R}^p)$ defined on a compact interval $I$. Let the probability distribution of $\textbf{Y}$ be $F_\textbf{Y}$. At each evaluation point, $t\in I$, $\textbf{Y}(t)$ is a $p$-variate vector with probability distribution $F_{\textbf{Y}(t)}$. The directional outlyingness for multivariate data is defined as:

\begin{equation}
\textbf{O}(\textbf{Y}, F_\textbf{Y}) = o(\textbf{Y}, F_\textbf{Y})\cdot \textbf{v},
\end{equation}
where $o(\textbf{Y}, F_\textbf{Y})$ is the outlyingness of $\textbf{Y}$ w.r.t. to $F_\textbf{Y}$ and $\textbf{v}$ is the spatial depth defined at point $t$ by
\begin{equation}
\textbf{v}(t) = \frac{|\textbf{Y}(t) - \textbf{Z}(t)|}{||\textbf{Y}(t) - \textbf{Z}(t)||},
\end{equation}
with $\textbf{Z}(t)$ being the unique median of $\textbf{Y}(t)$ w.r.t. $F_{\textbf{Y}(t)}$ (deepest point of $F_{\textbf{Y}(t)}$). $\textbf{v}(t)$ is a unit vector pointing from $\textbf{Z}(t)$ to $\textbf{Y}(t)$. \cite{dai2019directional} recommends  using a distance-based outlyingness measure, like the Stahel-Donoho outlyingness defined by:
\begin{equation}
\mathit{SDO}(\textbf{Y}(t), F_{\textbf{Y}(t)}) = \sup\limits_{||\textbf{u}||=1} \frac{||\textbf{u}^\top \textbf{Y}(t) - \text{median}(\textbf{u}^\top \textbf{Y}(t))||}{\text{mad}(\textbf{u}^\top \textbf{Y}(t))}.
\end{equation}

Thus, the Stahel-Donoho type directional outlyingness is given by:
\begin{equation}
\textbf{O}(\textbf{Y}, F_\textbf{Y}) = \mathit{SDO}(\textbf{Y}, F_\textbf{Y})\cdot \textbf{v}.
\end{equation}

Then the functional directional outlyingness (FO) is defined, to capture the \textit{overall} outlyingness for functional data, as:
\begin{equation}
\text{FO}(\textbf{Y}, F_\textbf{Y}) = \int_I ||\textbf{O}(\textbf{Y}(t), F_{\textbf{Y}(t)})||^2 w(t)dt,
\end{equation}
where $w(t)$ is a weight function defined on $I$. The mean directional outlyingness (\textbf{MO}) and variation of directional outlyingness (VO) were defined as :

\begin{equation}\textbf{MO}(\textbf{Y}, F_\textbf{Y}) = \int_I \textbf{O}(\textbf{Y}(t), F_{\textbf{Y}(t)}) w(t) dt, \end{equation}
and
\begin{equation}
\text{VO}(\textbf{Y}, F_\textbf{Y}) = \int_I ||\textbf{O}(\textbf{Y}(t), F_{\textbf{Y}(t)}) -  \textbf{MO}(\textbf{Y}, F_\textbf{Y})||^2 w(t)dt.
\end{equation}
These quantities measure the magnitude outlyingness and shape outlyingness of a functional observation, respectively. \cite{dai2019directional} further showed the relationship:
\begin{equation}
\text{FO}(\textbf{Y}, F_\textbf{Y}) = ||\textbf{MO}(\textbf{Y}, F_\textbf{Y})||^2 + \text{VO}(\textbf{Y}, F_\textbf{Y}),
\end{equation}
which decomposes the total functional outlyingness into the magnitude outlyingness and
the shape outlyingness.

In practice, the functional observations are obseved at a finite number of points, say $k$, on $I$, i.e., at points $t_1, t_2, \ldots, t_k \in I$. The finite dimensional version of $\textbf{MO}(\textbf{Y}, F_\textbf{Y})$ is defined as:
\begin{equation}
\textbf{MO}_k(\textbf{Y}, F_\textbf{Y}) = \frac{1}{k}\sum\limits_{i = 1} ^ k \textbf{O}(\textbf{Y}(t_i), F_{\textbf{Y}(t_i)})w(t_i),
\end{equation}
and the finite dimensional version of VO can defined in a similar manner.

After obtaining the $\textbf{MO}$ and VO for each curve, MS-Plot is then  a scatterplot of the points $(\textbf{MO}^\top, \text{VO})^\top$.  To detect outliers, a multivariate data whose columns are the \textbf{MO}s and VOs is formed, and a robust Mahalanobis distance is computed for each of the $(\textbf{MO}^\top, \text{VO})^\top$ pair in this data. The robust covariance matrix is estimated using the minimum covariate determinant (MCD) estimator \citep{rousseeuw1999fast}. The distribution of these robust distances is approximated using the F distribution \citep{hardinrocke3005}. Any observation with a robust distance greater than the cutoff obtained from the tails of the F distribution is flagged as an outlier.

The directional outlyingness and MS-Plot methods procedures are implemented mainly through the \code{dir_out()} and \code{msplot()} functions in \pkg{fdaoutlier}. These functions accept a matrix or data frame of dimension $n \times p$ for a univariate functional data, or an array of dimension $n \times p \times d$ for multivariate functional data (where $n$ is the number of functions/curves, $p$ is the number of evaluation points in the domain, and $d$ is the dimension of the functional data with, $d\ge 2$ for multivariate functional data). The \code{dir_out()} function computes the directional outlyingness matrix $\textbf{O}(\textbf{Y}, F_\textbf{Y})$, the mean directional outlyingness $\textbf{MO}(\textbf{Y}, F_\textbf{Y})$ and the variation of directional $\text{VO}(\textbf{Y}, F_\textbf{Y})$, while the \code{msplot()} function finds outliers using the mean and variation of outlyingness with the F approximation.

We illustrate identifying outliers with \code{msplot()} using \code{simulation_model5()} to generate data of 100 curves, out of which 10 are shape outliers with a different covariance structure.  The generated curves are observed on 50 domain points over the interval $[0, 1]$. A call to \code{simulation_model5()} returns a list containing the matrix of generated data and a vector containing the indices of the true outliers.
\begin{Schunk}
\begin{Sinput}
R> simdata <- simulation_model5(n = 100, p = 50,
+                               outlier_rate = 0.1, seed = 2)
R> dt <- simdata$data
R> dim(dt)
\end{Sinput}
\begin{Soutput}
[1] 100  50
\end{Soutput}
\begin{Sinput}
R> simdata$true_outliers
\end{Sinput}
\begin{Soutput}
 [1]  6 10 20 21 34 38 48 49 66 93
\end{Soutput}
\end{Schunk}
Next we pass the generated data to the \code{msplot()} function to detect the outliers in \code{dt}. By default, \code{msplot()} also produces a plot of the VO against the MO (or $\lvert\lvert \textbf{MO} \rvert\rvert$ in the case of a multivariate functional data) and returns a list containing \code{outliers} which is a vector of the indices of detected outliers. The plotting function can be turned off by setting the parameter \code{plot = FALSE}.
\begin{Schunk}
\begin{Sinput}
R> ms <- msplot(dts = dt, return_mvdir = T, plot = FALSE)
R> ms$outliers
\end{Sinput}
\begin{Soutput}
 [1]   6  10  20  21  34  38  48  49  51  66  93 100
\end{Soutput}
\end{Schunk}
Setting the additional parameter \code{return_mvdir = TRUE} ensures that vectors
of the mean and variation of outlyingness (MO and VO) of each curve are returned by \code{msplot()} (a matrix is returned for \textbf{MO} in the case of a multivariate functional data).
\begin{Schunk}
\begin{Sinput}
R> head(ms$mean_outlyingness)
\end{Sinput}
\begin{Soutput}
[1]  0.04718408 -0.76134612  1.30502807  0.20414153  0.81537363
[6]  2.27956644
\end{Soutput}
\begin{Sinput}
R> head(ms$var_outlyingness)
\end{Sinput}
\begin{Soutput}
[1] 0.2115806 0.1777422 0.2015300 0.3898691 0.1550651 2.2730378
\end{Soutput}
\end{Schunk}
\begin{figure}[t!]
\centering
\includegraphics{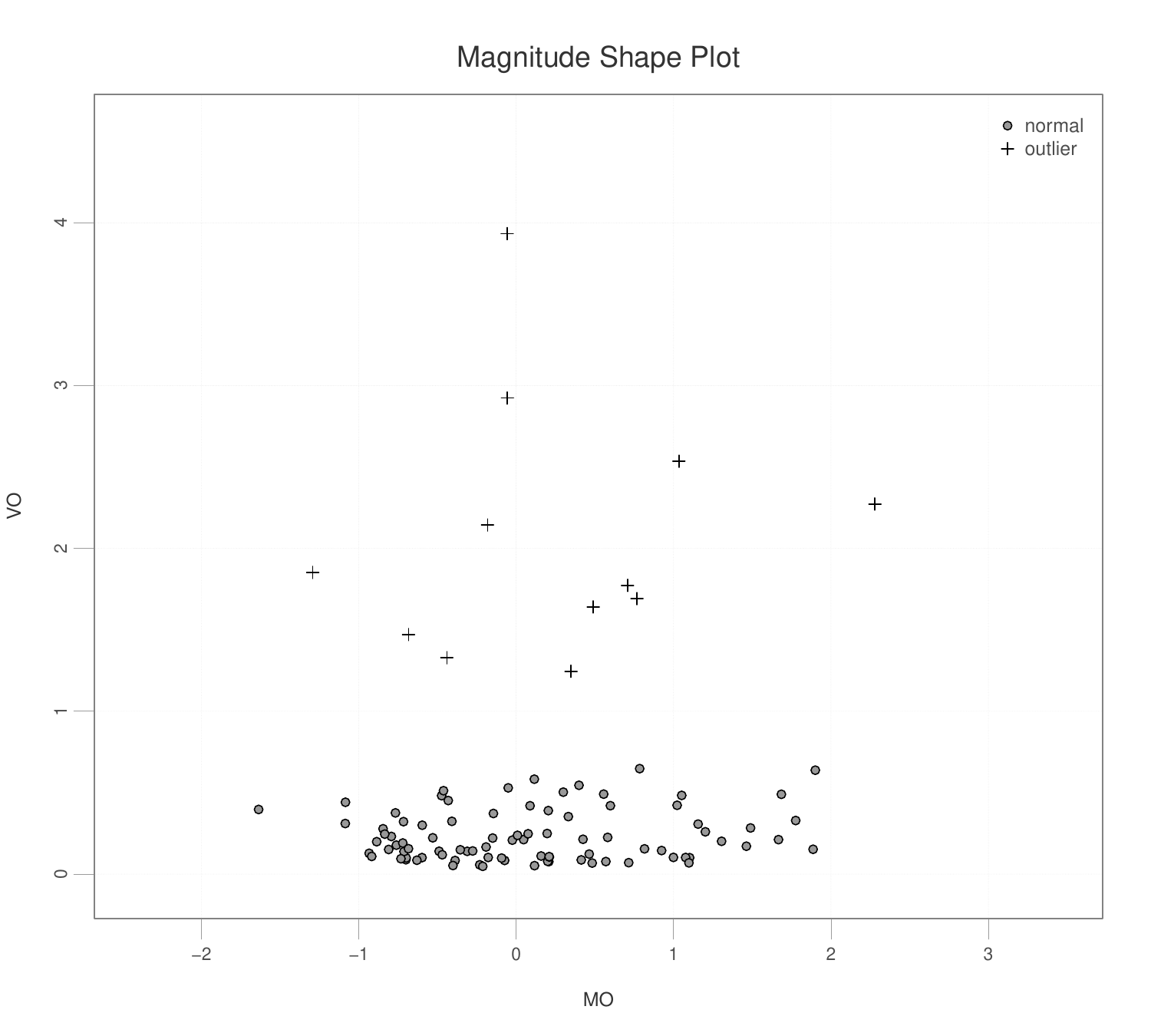}
\caption{\label{fig:msplot} MS-Plot: Plot of the VO against the MO.}
\end{figure}

The MS-Plot produced by the \code{msplot()} function when the parameter \code{plot = TRUE} is shown in Figure~\ref{fig:msplot}. Additional parameters \code{plot_title}, \code{title_cex}, \code{show_legend}, \code{ylabel} and \code{xlabel} can be passed to \code{msplot()} to further customise the MS-Plot generated.

If the aim is to compute either the MO and VO or the directional outlyingness matrix  (or array for multivariate functional data), without the  need for identifying outliers, then the \code{dir_out()} function, which is called by \code{msplot()} can be used directly. This returns a list containing the mean and variance of outlyingness, and the directional outlyingness matrix (if the parameter \code{return_dir_matrix = TRUE}).
\begin{Schunk}
\begin{Sinput}
R> simdir <- dir_out(dts = dt, return_dir_matrix = T)
R> head(simdir$mean_outlyingness)
\end{Sinput}
\begin{Soutput}
[1]  0.04718408 -0.76134612  1.30502807  0.20414153  0.81537363
[6]  2.27956644
\end{Soutput}
\begin{Sinput}
R> head(simdir$var_outlyingness)
\end{Sinput}
\begin{Soutput}
[1] 0.2115806 0.1777422 0.2015300 0.3898691 0.1550651 2.2730378
\end{Soutput}
\begin{Sinput}
R> dim(simdir$dirout_matrix)
\end{Sinput}
\begin{Soutput}
[1] 100  50
\end{Soutput}
\end{Schunk}
\subsection{Total variation depth and modified shape similarity index}
Suppose $Y:I \rightarrow \mathbb{R}$ is a stochastic process defined on the interval $I$ in $\mathbb{R}$. Let the distribution of $Y$ be $F_Y$. For a function $y$, let $R_y(t)$ be the indicator function:

\begin{equation}
R_y(t) = \mathbbm{1}\{Y(t) \le y(t)\},
\end{equation}

for $t \in I$. The functional total variation depth \citep{huang2019decomposition} of the function $y$ w.r.t. $F_Y$ is then defined as:

\begin{equation}
\mathit{TVD}(y, F_Y) = \int_{I} D_{y}(t) w(t) dt,\end{equation}
where $w(t)$ is a weight function and $D_{y}(t)$ is the pointwise total variation depth given by:
\begin{equation}
D_{y}(t) = \textrm{var}\{R_y(t)\} = \mathbb{P}\{Y(t) \le y(t)\}\mathbb{P}\{Y(t) > y(t)\}.
\end{equation}
The constant weight function $w(t) = \frac{1}{|I|}$ is suggested in \citep{huang2019decomposition} but other weight functions (that place more emphasis on different regions of the interval) can be used in the formulation of the functional total variation depth. The pointwise total variation depth $D_y(t)$ can be decomposed into a shape and magnitude component by breaking up the variance $\textrm{var}\{R_y(t)\}$ using the law of total variance:

\begin{equation}
D_y(t) = \textrm{var}\{R_y(t)\} = \textrm{var}[\mathbb{E}\{R_y(t)| R_y(s)\}] + \mathbb{E}[\textrm{var}\{R_y(t) | R_y(s)\}],
\end{equation}

for $s, t \in I$ and $s = t - \Delta$. The shape similarity index of the functional observation $y$ in a given time span $\Delta$ is then the weighted ratio of the shape component $\textrm{var}[\mathbb{E}\{R_y(t)| R_y(s)\}]$ to the total variation depth over the interval $I$:

\begin{equation}
\mathit{SS}(y, \Delta) = \int_I S_y(t, \Delta) v(t, \Delta) dt,
\end{equation}

where
$$S_y(t, \Delta) = \begin{cases}
      \textrm{var}[\mathbb{E}\{R_y(t)| R_y(s)\}]/D_y(t) & D_y(t) \neq 0 \\
     1 & D_y(t) = 0, \\
   \end{cases}
$$
and the weight function $v(t, \Delta)$ is the normalised changes in $y(t)$ over the interval $I$:

$$v(t, \Delta) = \frac{|y(t) - y(t - \Delta)|}{\int_I |y(t) - y(t - \Delta)|}.$$

The shape similarity index is a measure of shape outlyingness with small indices associated with shape outliers. However, when $D_y(t)$ is very small, the shape similarity index may not be small enough, so \cite{huang2019decomposition} further defined the modified shape similarity index (MSS) by shifting $(y(t - \Delta), y(t))$ to the centre. The modified shape similarity index is defined as:

\begin{equation}
\mathit{MSS}(y, \Delta) = \int_I S_{\tilde{y}} (t, \Delta) v(t, \Delta)dt,
\end{equation}

where $\widetilde{y}$ is given by:
$$\widetilde{y}(s, \Delta) = \begin{cases}
\textrm{median}\{Y(s)\} & s = t \\
f(s) - f(s + \Delta) + \textrm{median}[Y(s + \Delta)] & s = t - \Delta,
\end{cases} $$

and

$$S_{\tilde{y}} (t, \Delta) = \textrm{var}(\mathbb{E}[R_{\tilde{y}}(t)| R_{\tilde{y}}(s)])/D_{\tilde{y}}(t)\}.$$

Details of the empirical versions of the total variation depth, the shape similarity index and its modified version are presented in the Appendix of \cite{huang2019decomposition}. The total variation depth and the modified similarity index are implemented in the \code{total_variation_depth()} function of \pkg{fdaoutlier} using \proglang{C++} through \proglang{R}'s \code{.Call} interface for faster and efficient computation. This function accepts only a matrix, and calling it suffices to compute both the total variation depth and the modified shape similarity index. \code{total_variation_depth()} returns a list containing both the total variation depth and the modified shape similarity index:
\begin{Schunk}
\begin{Sinput}
R> tvdepth <- total_variation_depth(dt)
R> head(tvdepth$mss)
\end{Sinput}
\begin{Soutput}
[1] 0.6388217 0.6208659 0.6975233 0.6853633 0.6618313 0.2167154
\end{Soutput}
\begin{Sinput}
R> head(tvdepth$tvd)
\end{Sinput}
\begin{Soutput}
[1] 0.224578 0.156484 0.114386 0.197932 0.178926 0.072398
\end{Soutput}
\end{Schunk}
In order to identify outliers, shape outliers are first identified and removed using a classical boxplot on the modified shape similarity indices. A functional boxplot is then used on the remaining curves (to identify magnitude outliers) using the total variation depth to identify their $50\%$ central region (relative to the original number of curves).  The \code{tvdmss()} function implements these steps to detect the magnitude and shape outliers. It returns a function containing the indices of the magnitude outliers, shape outliers, and the combined (shape and magnitude) outliers. This is illustrated using the generated data \code{dt} from the previous section.
\begin{Schunk}
\begin{Sinput}
R> tvoutlier <- tvdmss(dts = dt)
R> tvoutlier$shape_outliers
\end{Sinput}
\begin{Soutput}
 [1]  6 10 20 21 34 38 48 49 66 93
\end{Soutput}
\begin{Sinput}
R> tvoutlier$magnitude_outliers
\end{Sinput}
\begin{Soutput}
NULL
\end{Soutput}
\begin{Sinput}
R> tvoutlier$outliers
\end{Sinput}
\begin{Soutput}
 [1]  6 10 20 21 34 38 48 49 66 93
\end{Soutput}
\end{Schunk}
In this case the total variation depth identifies all the shape outliers correctly and does not detect any magnitude outliers when compared to the index of the true outliers of the generated data:

\begin{Schunk}
\begin{Sinput}
R> simdata$true_outliers
\end{Sinput}
\begin{Soutput}
 [1]  6 10 20 21 34 38 48 49 66 93
\end{Soutput}
\end{Schunk}

Additional arguments can be passed to the parameters \code{emp_factor_mss}, \code{emp_factor_tvd}, and \code{central_region_tvd} of  \code{tvdmss()} to control the classical boxplot of the modified shape similarity index and the functional boxplot of the total variation depth.

\subsection{Outlier detection using sequential transformations}

\cite{dai2020sequential} proposed using some sequence of tranformations on the functional data to identify and classify functional outliers. By transforming the functional data, it is possible to turn shape outliers into magnitude outliers, consequently making it easier to identify shape outliers. More formally, let $\{Y_i\}_{i = 1}^n$ be a set of functional observations in the space of continous functions $C(I)$ defined on an interval $I \in \mathbb{R}$. Suppose that $Y_i \sim F_Y$, and let $\Gamma$ be a transformation that is also defined on $C(I)$. Furthermore, let $F_{\Gamma(Y)}$ be the distribution of the transformed data $\{\Gamma(Y_i)\}_{i = 1}^n$. \cite{dai2020sequential} proposed the following  algorithm for functional outlier detection and taxonomy.

\begin{figure}[ht]

\begin{algorithm}[H]
\SetAlgoLined
Identify from $\{Y_i\}_{i = 1}^n$ the magnitude outliers using the functional boxplot, and denote the set of identified outliers by $S_0$. These are the $\Gamma_0-$outliers (magnitude outliers). \\
Apply transformation $\Gamma_1$ on $\{Y_i\}_{i = 1}^n$ to get  $\{\Gamma_1(Y_i)\}_{i = 1}^n$. \\
Repeat step 1 on $\{\Gamma_1(Y_i)\}_{i = 1}^n$ to obtain the set of detected outliers $S_1$; $S_1\setminus S_0$ are the $\Gamma_1-$shape outliers. \\
Apply transformation $\Gamma_2$ on $\{\Gamma_1(Y_i)\}_{i = 1}^n$ to get  $\{\Gamma_2 \circ \Gamma_1(Y_i)\}_{i = 1}^n$. \\

Repeat step 1 on $\{\Gamma_2 \circ \Gamma_1(Y_i)\}_{i = 1}^n$ to obtain the set of detected outliers $S_2$; $S_2\setminus S_1\cup S_0$ are the $\Gamma_2 \circ \Gamma_1-$shape outliers. \\

Repeat steps 4 and 5 if more transformations are considered. \\ \caption{Functional outlier detection using sequential transformations.}\label{alg:seq_transformation}
\end{algorithm}
\end{figure}

\cite{dai2020sequential} proposed the following useful (sequence of) transformations to identify and classify outliers:

\textit{Shifting and normalization of Curves:}  $\mathcal{T}_2 \circ \mathcal{T}_1 \circ \mathcal{T}_0 (Y_i)$

This sequence involves first identifying the magnitude outliers using functional boxplot. This is the $\mathcal{T}_0$ transformation and the identified outliers are the $\mathcal{T}_0$~outliers (magnitude outliers). The second transformation $\mathcal{T}_1$ involves shifting the raw curves $Y_i$ to their centres:

\begin{equation}
\mathcal{T}_1(Y)(t) = Y(t) -  \lambda(I)^{-1}\int_I Y(t)dt,
\end{equation}
where $\lambda(I)$ is the Lebesgue measure of the interval $I$. The $\mathcal{T}_1$~outliers are then identified using functional boxplot (step 3 of Algorithm 1). The third transformation $\mathcal{T}_2$ involves normalizing the centered curves, i.e., $\{\mathcal{T}_1(Y_i)\}_{i = 1}^n$, with their $L_2$ norms:

\begin{equation}
\mathcal{T}_2(Y)(t) = \frac{\mathcal{T}_1(Y)(t)}{\left[ \int_I \{\mathcal{T}_1(Y)(t)\}^2 dt \right]^{1/2}}.
\end{equation}

\textit{Derivatives of curves:} $\mathcal{D}_2 \circ \mathcal{D}_1 \circ \mathcal{D}_0 (Y_i)$

The $\mathcal{D}_0$ transformation first involves identifying the magnitude outliers using a functional boxplot without transforming the data (same as $\mathcal{T}_0$). These are the $\mathcal{D}_0$~outliers. The second transformation $\mathcal{D}_1$ involves finding the derivative of the curves, and the third transformation $\mathcal{D}_2$ computes the derivative of $\mathcal{D}_1(Y_i)$  again. After each transformation, outliers are identified using functional boxplot as indicated in Algorithm 1. $\mathcal{D}_1$ and $\mathcal{D}_2$ transforms are implemented in \pkg{fdaoutlier} by differencing the observed points of the functions on the domain.

\textit{Directional outlyingness:}   $\mathcal{O}(\textbf{Y}(t))$

For multivariate functional data $\{\textbf{Y}_i\}_{i = 1}^n$ taking values in $\mathbb{R}^d$, the directional outlyingness transformation $\mathcal{O}$ is especially useful. This transformation changes the multivariate functional observation $\textbf{Y}_i$ to univariate functional data $Y_i$ by finding the pointwise directional outlyingness described in Section~\ref{subsec:2.1} \citep{dai2019directional}. The univariate functional data (of the outlyingness values) can then be investigated for outliers, e.g., using functional boxplot with a one-sided ordering like the (one-sided) extreme rank length depth \citep[see][and \citeauthor{dai2020sequential} \citeyear{dai2020sequential}]{mari2017}.

Other transformations and sequences suggested in \cite{dai2020sequential} include elimination of phase variations using a warping function:
\begin{equation}
\mathcal{R}(Y)(t) = Y(r(t)),
\end{equation}

where $r(t)$ is a warping function on $I$. Eliminating phase variations using $\mathcal{R}(Y)$ may make it easier to detect shape outliers. Other possible sequences of transformations are: $\mathcal{D}_1 \circ \mathcal{T}_1 \circ \mathcal{T}_0(Y)$ and $\mathcal{D}_2 \circ \mathcal{D}_1 \circ \mathcal{T}_2 \circ \mathcal{T}_1 \circ \mathcal{T}_0(Y)$.

In the intermediate steps of identifying outliers using functional boxplots, possible depths and outlyingness measures that can be used to order the functions are: modified band depth of \cite{Romo_depths} (MBD), $j^{th}$ order integrated depth of \cite{nagy07depth} $(FD_j)$, the $L^{\infty}$ depth \citep{long2015}, and extreme rank length depth (ERLD) of \citep{mari2017}. Other methods include the robust Mahalanobis distance (RMD) of the $(\textbf{MO}^\top, \text{VO})^\top$ pair, obtained from the directional outlyingness in Section~\ref{subsec:2.1}, and directional quantile (DQ) \citep{mari2017}. DQ, RMD, and $L^{\infty}$ are distance-based, while MBD, $FD_j$, and ERLD are based on ranks. \cite{dai2020sequential} suggested using the distance-based methods, especially when the number of evaluation points on the interval $I$ is small, as rank-based methods might suffer from a large number of ties. The distance-based methods also achieved the best results for detecting shape outliers in the simulation tests consisting of various shape outliers conducted in \cite{dai2020sequential}. However, some transformations may require the use of specific ordering methods, e.g., the one-sided ERLD is best used with the $\mathcal{O}$ transformation since it generates univariate functional data made up of point-wise directional outlyingness, and we want to consider only large values of these curves as extremes (rather than use a typical functional depth like MBD which considers both small and large values of curves as extremes).  The \pkg{fdaoutlier} package implements all the transformations mentioned in \cite{dai2020sequential} except for the $\mathcal{R}(Y)(t)$ transformation which involves the use of a warping function. The ordering measures: band depth (BD) and MBD, $L^{\infty}$, $DQ$, RMD, TVD, and ERLD (both one and two-sided) are available in \pkg{fdaoutlier} for ordering the functions in the intermediate functional boxplots.

The \code{seq_transform()} function in \pkg{fdaoutlier} finds outliers using sequential transformations. Like the other functions in \pkg{fdaoutlier}, \code{seq_transform()} accepts a matrix or data frame (of size $n$ observations by $p$ evaluation points) for a univariate functional data and an array (of size $n$ observation by $p$ evaluation points by $d$ dimension). The sequence of transformations to apply on the data is specified to the \code{sequence} parameter which accepts a character vector containing a combination of the following strings: \code{"T0"}, \code{"D0"}, \code{"T1"}, \code{"T2"}, \code{"D1"}, \code{"D2"}, and \code{"O"}. The strings \code{"T0"}, \code{"T1"} and \code{"T2"} represent the tranformations $\mathcal{T}_0$, $\mathcal{T}_1$ and $\mathcal{T}_2$ respectively, while the strings \code{"D0"}, \code{"D1"}, and \code{"D2"} represent $\mathcal{D}_0$, $\mathcal{D}_1$ and $\mathcal{D}_2$ respectively. The string \code{"O"} indicates the outlyingness transformation $\mathcal{O}(\textbf{Y})(t)$. Thus, to specify the sequence of tranformations: $\mathcal{D}_1 \circ \mathcal{T}_1 \circ \mathcal{T}_0(Y)$, one should pass the argument \code{c("T0", "T1", "D1")} to the parameter \code{sequence} in the call to \code{seq_tranform()}, i.e., set \code{sequence = c("T0", "T1", "D1")}. We provide some examples below on the use of the \code{seq_transform()} function for detecting outliers using some suggested sequences in \cite{dai2020sequential}.

First we generate some data with outliers from \code{simulation_model4()}:
\begin{Schunk}
\begin{Sinput}
R> simdata4 <- simulation_model4(n = 100, p = 50, outlier_rate = 0.05,
+                                deterministic = T, seed = 50)
R> dt4 <- simdata4$data
\end{Sinput}
\end{Schunk}
Next, we call the \code{seq_transform()} using the sequence $\mathcal{T}_2 \circ \mathcal{T}_1 \circ \mathcal{T}_0(Y)(t)$ and specifying modified band depth as the ordering function of choice for the intermediate functional boxplots.
\begin{Schunk}
\begin{Sinput}
R> seq1 <- seq_transform(dts = dt4,
+                        sequence = c("T0", "T1", "T2"),
+                        depth_method = "mbd")
R> seq1$outliers
\end{Sinput}
\begin{Soutput}
$T0
integer(0)

$T1
[1] 43 53

$T2
[1] 43 53 96
\end{Soutput}
\end{Schunk}
\code{seq_transform()} returns a list of named lists, with each named lists containing the indices of the outliers found at each step of the sequence of transformations. The names of the lists are the different transformations conducted at each step. In this example, the sequence $\mathcal{T}_2 \circ \mathcal{T}_1 \circ \mathcal{T}_0(Y)(t)$ (with modified band depth) identifies only three of the five true outliers contained in the simulated data:
\begin{Schunk}
\begin{Sinput}
R> unique(unlist(seq1$outliers))
\end{Sinput}
\begin{Soutput}
[1] 43 53 96
\end{Soutput}
\begin{Sinput}
R> simdata4$true_outliers
\end{Sinput}
\begin{Soutput}
[1] 20 43 53 70 96
\end{Soutput}
\end{Schunk}
Next we try the sequence $\mathcal{D}_1 \circ \mathcal{T}_1 \circ \mathcal{T}_0(Y)(t)$ but now with the total variation depth in the intermediate functional boxplot:
\begin{Schunk}
\begin{Sinput}
R> seq2 <- seq_transform(dts = dt4,
+                        sequence = c("T0", "T1", "D1"),
+                        depth_method = "tvd")
R> seq2$outliers
\end{Sinput}
\begin{Soutput}
$T0
integer(0)

$T1
[1] 43 53

$D1
integer(0)
\end{Soutput}
\end{Schunk}
The sequence $\mathcal{D}_1 \circ \mathcal{T}_1 \circ \mathcal{T}_0(Y)(t)$ with total
variation depth identifies only two of the five true outliers as seen below:
\begin{Schunk}
\begin{Sinput}
R> unique(unlist(seq2$outliers))
\end{Sinput}
\begin{Soutput}
[1] 43 53
\end{Soutput}
\begin{Sinput}
R> simdata4$true_outliers
\end{Sinput}
\begin{Soutput}
[1] 20 43 53 70 96
\end{Soutput}
\end{Schunk}

Another suggested sequence is the sequence $\mathcal{D}_2 \circ \mathcal{D}_1 \circ \mathcal{D}_0(Y)(t)$. We use this sequence but now with the $L^\infty$ depth as the
ordering function:
\begin{Schunk}
\begin{Sinput}
R> seq3 <- seq_transform(dts = dt4,
+                        sequence = c("D0", "D1", "D2"),
+                        depth_method = "linfinity")
R> seq3$outliers
\end{Sinput}
\begin{Soutput}
$D0
integer(0)

$D1
integer(0)

$D2
integer(0)
\end{Soutput}
\end{Schunk}
This time, the sequence $\mathcal{D}_2 \circ \mathcal{D}_1 \circ \mathcal{D}_0(Y)(t)$ does not identify any of the outliers. This is not surprising as the sequence  $\mathcal{D}_2 \circ \mathcal{D}_1 \circ \mathcal{D}_0(Y)(t)$ is advisable for identifying pure magnitude (captured by $\mathcal{D}_0)$ and pure shape outliers (captured by $\mathcal{D}_1$ and $\mathcal{D}_2$) and this result is in line with the results of the simulation tests conducted in \cite{dai2020sequential} where the sequence $\mathcal{D}_2 \circ \mathcal{D}_1 \circ \mathcal{T}_0(Y)(t)$ performed the worst on this simulation model (See Table 4 of \cite{dai2020sequential}). Note that the sequence $\mathcal{D}_2 \circ \mathcal{D}_1 \circ \mathcal{D}_0(Y)(t)$ can also be specified with the \code{sequence} argument \code{c("D0", "D1", "D1")} or \code{c("D0", "D2", "D2")} or \code{c("T0", "D1", "D2")} since both \code{"D1"} and \code{"D2"} do the same thing, i.e., perform a lag-1 differencing and \code{"D0"} and \code{"T0"} also do the same thing (identify magnitude outliers in the raw untransformed data). When there are repeated transformations in the argument passed to \code{sequence} (e.g., when \code{sequence = c("D0", "D1", "D1")} is passed), a warning is shown  and the labels of the output outliers list are changed, so that outliers for the two $\mathcal{D}_1$ transformations are accessed with \code{output$outliers$D1_1} and \code{output$outliers$D1_2} respectively:
\begin{Schunk}
\begin{Sinput}
R> seq4 <- seq_transform(dt = dt4,
+                        sequence = c("D0", "D1", "D1"),
+                        depth_method = "linfinity")
R> seq4$outliers
\end{Sinput}
\begin{Soutput}
$D0
integer(0)

$D1_1
integer(0)

$D1_2
integer(0)
\end{Soutput}
\end{Schunk}
Sometimes, it may be necessay to inspect or save the intermediate transformed data for further analysis. Each intermediate transformed data can be saved by setting the parameter \code{save_data = TRUE} in the call to \code{seq_tranform()}. These data can then be assessed with the form \code{object_name$transformed_data$transform}:
\begin{Schunk}
\begin{Sinput}
R> seq5 <- seq_transform(dt = dt4,
+                        sequence = c("D0", "D1", "D1"),
+                        depth_method = "mbd", save_data = T)
R> str(seq5$transformed_data$D1_1)
\end{Sinput}
\begin{Soutput}
 num [1:100, 1:49] 0.409 0.659 0.657 0.307 0.397 ...
 - attr(*, "dimnames")=List of 2
  ..$ : NULL
  ..$ : chr [1:49] "2" "3" "4" "5" ...
\end{Soutput}
\end{Schunk}
As mentioned earlier, the $\mathcal{O}(\textbf{Y})(t)$ transformation should be used with a one-sided ERLD ordering (in the functional boxplot) so that only large values of the resulting outlyingness data are considered as extremes:
\begin{Schunk}
\begin{Sinput}
R> seq6 <- seq_transform(dt = dt4,
+                        sequence = "O",
+                        depth_method = "erld",
+                        erld_type = "one_sided_right")
R> seq6$outliers
\end{Sinput}
\begin{Soutput}
$O
[1] 18 43 53 70
\end{Soutput}
\end{Schunk}
The additional parameter \code{erld_type} specifies whether large values should be considered as extremes (\code{erld_type = "one_sided_right"}), or small values should be considered as extremes (\code{erld_type = "one_sided_left"}) or both small and large values should be considered as extremes (\code{erld_type = "two_sided"}). The two sided ordering is used by default if \code{erld_type} is not specified.

\subsection{Massive unsupervised outlier detection}
The massive unsupervised outlier detection (MUOD) detects and classifies outliers into shape, magnitude, and amplitude outliers. It was proposed in \cite{azcorra2018unsupervised} as a support method to identify influential users in social networks. MUOD works by computing for each curve three indices which measure outlyingness in terms of shape, magnitude, and amplitude. The shape index of $Y_i$ w.r.t. $F_Y$ denoted by $I_S(Y_i, F_Y)$ is defined as

\begin{equation}
I_S(Y_i, F_Y) = \left|\frac{1}{n}\sum\limits_{j=1}^n \hat{\rho}(Y_i, Y_j) - 1 \right|,
\end{equation}

where $\hat{\rho}(Y_i, Y_j)$ is the Pearson correlation coefficient between the observed points of curves $Y_i$ and $Y_j$, given by

\begin{equation}
\hat{\rho}(Y_i, Y_j) = \frac{\text{cov}(Y_i, Y_j)}{s_{Y_i}s_{Y_j} },\ \ \ \ \ \ s_{Y_i}, s_{Y_j} \ne 0.
\end{equation} The magnitude and amplitude indices of $Y_i$ w.r.t. $F_Y$ are defined are:

\begin{equation}I_M(Y_i, F_Y) = \left|\frac{1}{n}\sum\limits_{j = 1}^n\hat{\alpha}_j \right|,
\end{equation}
and
\begin{equation}
I_A(Y_i, F_Y) = \left|\frac{1}{n}\sum\limits_{j = 1}^n\hat{\beta}_j -1 \right|,
\end{equation}
respectively, where
$$
\hat{\beta}_j = \frac{\text{cov}(Y_i, Y_j)}{s_{Y_j}^2}, \ \ \ \ \ \ s_{Y_j}^2 \ne 0,
$$

$$\hat{\alpha}_j = \bar{x_i} - \hat{\beta}_j \bar{x}_j,$$

and

$$\bar{x_i} = \frac{\sum_{t \in \mathcal{I}}Y_i(t)}{p}.$$

Generally, magnitude outliers will have larger magnitude indices, and the same applies to shape and amplitude outliers. To identify a cutoff for the indices, \cite{azcorra2018unsupervised} proposed to use a \textit{``tangent''} method, which searches for the line tangent to the maximum index and uses as cutoff the point where this tangent line intercepts the x-axis. This method is especially problematic and prone to false positives, as pointed out by \cite{vinue2020robust}. A alternative is to use a classical boxplot on the indices to identify extremely large indices.

MUOD is implemented in \pkg{fdaoutlier} and can be accessed through the \code{muod()} function. The tangent method or the classical boxplot can be specified for determining the indices cutoffs. The function returns a list containing the outliers and the MUOD indices. The outliers list contains vectors with names \code{shape}, \code{amplitude} and \code{magnitude} all containing the indices of the detected shape, amplitude, and amplitude outliers, respectively.

\begin{Schunk}
\begin{Sinput}
R> simdata1 <- simulation_model1(n = 100, p = 50,
+                                outlier_rate = 0.1, seed = 2)
R> moutlier <- muod(dts = simdata1$data, cut_method = "tangent")
R> moutlier$outliers
\end{Sinput}
\begin{Soutput}
$shape
[1] 100  58  35  79  21  40  50  14

$amplitude
[1]  51 100  58  14  94

$magnitude
 [1] 20 21 66 48 38 34 10 49  6 93 67
\end{Soutput}
\begin{Sinput}
R> moutlier2 <- muod(dts = simdata1$data, cut_method = "boxplot")
R> moutlier2$outliers
\end{Sinput}
\begin{Soutput}
$shape
[1] 100  58  35  79  21  40

$amplitude
[1]  51 100  58

$magnitude
 [1] 20 21 66 48 38 34 10 49  6 93
\end{Soutput}
\end{Schunk}
Furthermore, the muod magnitude ($I_M$), amplitude ($I_A$) and shape ($I_S$) indices can be accessed for further analysis:
\begin{Schunk}
\begin{Sinput}
R> str(moutlier$indices)
\end{Sinput}
\begin{Soutput}
'data.frame':	100 obs. of  3 variables:
 $ shape    : num  0.1007 0.092 0.0845 0.1153 0.0721 ...
 $ magnitude: num  0.404 0.437 1.188 1.21 0.371 ...
 $ amplitude: num  0.08081 0.26492 0.00309 0.51515 0.11212 ...
\end{Soutput}
\end{Schunk}
%



\section{Usage examples} \label{sec:illustrations}

In this section, we demonstrate the use of the \pkg{fdaoutlier} package on some
common real datasets in the functional data analysis literature. In particular, we replicate some of the application examples from \cite{dai2018multivariate} and \cite{dai2020sequential}. First, we analyse the Spanish ('aemet') weather dataset, followed by the population growth dataset. These datasets have been analysed extensively in the literature, and they provide meaningful applications for outlier detection in functional data analysis.

The Spanish weather data contains the daily average temperature, log precipitation, and wind speed of 73 Spanish weather stations measured between 1980-2009. This data was analysed in \cite{dai2018multivariate}, \cite{dai2019directional}, and \cite{dai2020sequential} where the directional outlyingness, MS-Plot, and sequential transformation methods were proposed. The data is originally available in the \pkg{fda.usc} package (with the name \code{aemet}) but a stripped-down version is also made available in \pkg{fdaoutlier} (with the name \code{spanish_weather}). In this analysis, we focus on the average temperature and log precipitation, and the aim is to find outlying curves (weather stations with outlying weather data). Following \cite{dai2019directional}, we smooth the data with 11 B-spline basis functions by obtaining a smoothing matrix (without roughness penalty) using the \pkg{fda.usc} package.
\begin{Schunk}
\begin{Sinput}
R> library("fda.usc")
R> data("spanish_weather")
R> b_spline <- create.bspline.basis(c(0, 365), nbasis = 11)
R> smoothing_matrix <- S.basis(tt = 0.5:364.5, basis =  b_spline)
R> temp <- spanish_weather$temperature %*% smoothing_matrix
R> logprecip <- spanish_weather$log_precipitation %*% smoothing_matrix
\end{Sinput}
\end{Schunk}
\begin{figure}[t!]
\centering
\includegraphics{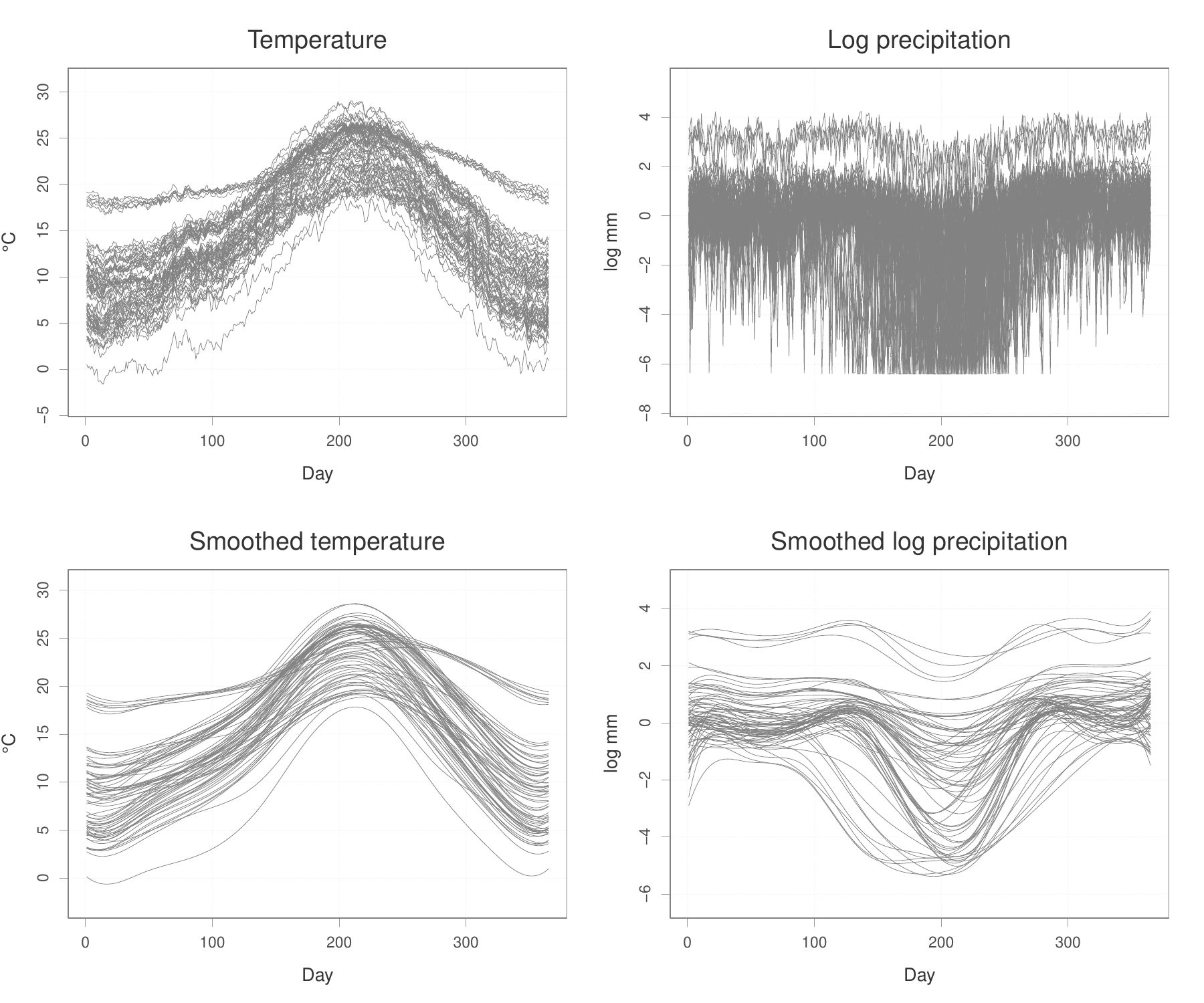}
\caption{\label{fig:aemet1} Plot of temperature and log precipitation and their smoothed (with 11 B-spline basis) versions.}
\end{figure}

\begin{figure}[t!]
\centering
\includegraphics{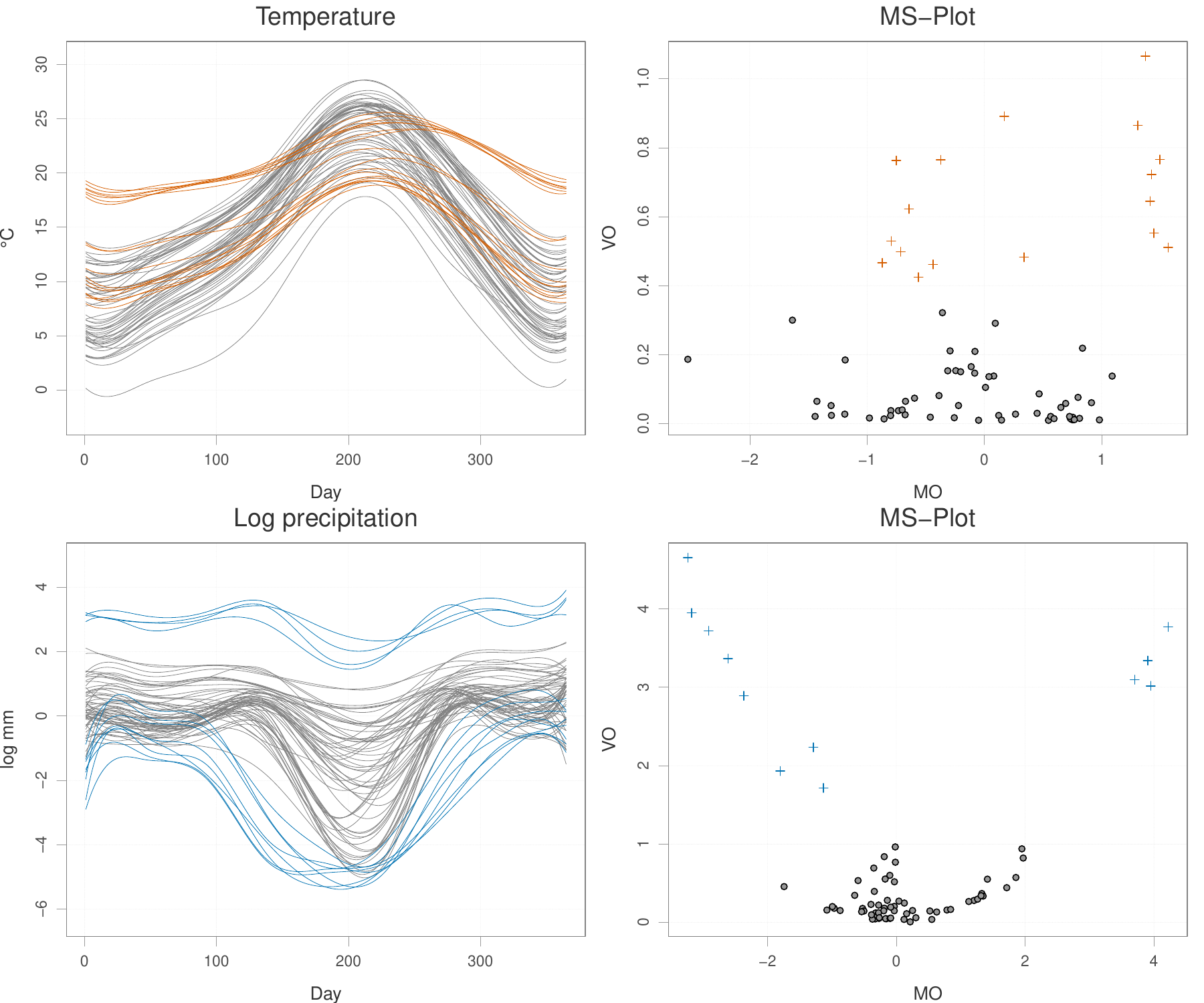}
\caption{\label{fig:aemet2} Plot of temperature and log precipitation and their MS-Plots. Lines and points in color are outliers.}
\end{figure}

A plot of the original and smoothed versions of the temperature and log precipitation data is shown in Figure \ref{fig:aemet1}. Next, we apply MS-Plot on the individual smoothed temperature and log precipitation data using the \code{msplot()} function to detect the marginal outliers.
\begin{Schunk}
\begin{Sinput}
R> temp_ms <- msplot(dts = temp, plot = F)
R> logprecip_ms <- msplot(dts = logprecip, plot = F)
\end{Sinput}
\end{Schunk}
Using the indices of the outliers returned, we can identify the weather stations detected as marginal outliers using the station information data (\code{station_info}).
\begin{Schunk}
\begin{Sinput}
R> head(spanish_weather$station_info$name[temp_ms$outliers])
\end{Sinput}
\begin{Soutput}
[1] "A CORUÑA"                        
[2] "A CORUÑA/ALVEDRO"                
[3] "SANTIAGO DE COMPOSTELA/LABACOLLA"
[4] "ASTURIAS/AVILÉS"                 
[5] "OVIEDO"                          
[6] "TARIFA"                          
\end{Soutput}
\begin{Sinput}
R> head(spanish_weather$station_info$name[logprecip_ms$outliers])
\end{Sinput}
\begin{Soutput}
[1] "LOGROÑO/AGONCILLO"               
[2] "FUERTEVENTURA/AEROPUERTO"        
[3] "LANZAROTE/AEROPUERTO"            
[4] "LAS PALMAS DE GRAN CANARIA/GANDO"
[5] "COLMENAR VIEJO/FAMET"            
[6] "MADRID/TORREJÓN"                 
\end{Soutput}
\end{Schunk}
Using the vectors of the mean and variation of directional outlyingness returned by \code{msplot()}, we can make a plot of the outliers detected and the plots of VO against MO as shown in Figure~\ref{fig:aemet2}.

We can also apply \code{msplot()} on the multivariate functional constructed by combining both the smoothed temperature and log precipitation using an \code{array} in order to detect and identify the joint temperature and log precipitation outliers:
\begin{Schunk}
\begin{Sinput}
R> joint_dt <- array(data = c(as.vector(temp), as.vector(logprecip)),
+                      dim = c(nrow(temp), ncol(temp), 2))
R> joint_ms <- msplot(joint_dt, plot = F)
R> joint_ms$outliers
\end{Sinput}
\begin{Soutput}
 [1]  1  2  3  9 20 21 31 33 34 35 36 39 44 52 55 56 57 58 59 60 66 70
\end{Soutput}
\begin{Sinput}
R> head(spanish_weather$station_info$name[joint_ms$outliers])
\end{Sinput}
\begin{Soutput}
[1] "A CORUÑA"                        
[2] "A CORUÑA/ALVEDRO"                
[3] "SANTIAGO DE COMPOSTELA/LABACOLLA"
[4] "ASTURIAS/AVILÉS"                 
[5] "TARIFA"                          
[6] "SANTANDER/PARAYAS"               
\end{Soutput}
\end{Schunk}

\begin{figure}[t!]
\centering
\includegraphics{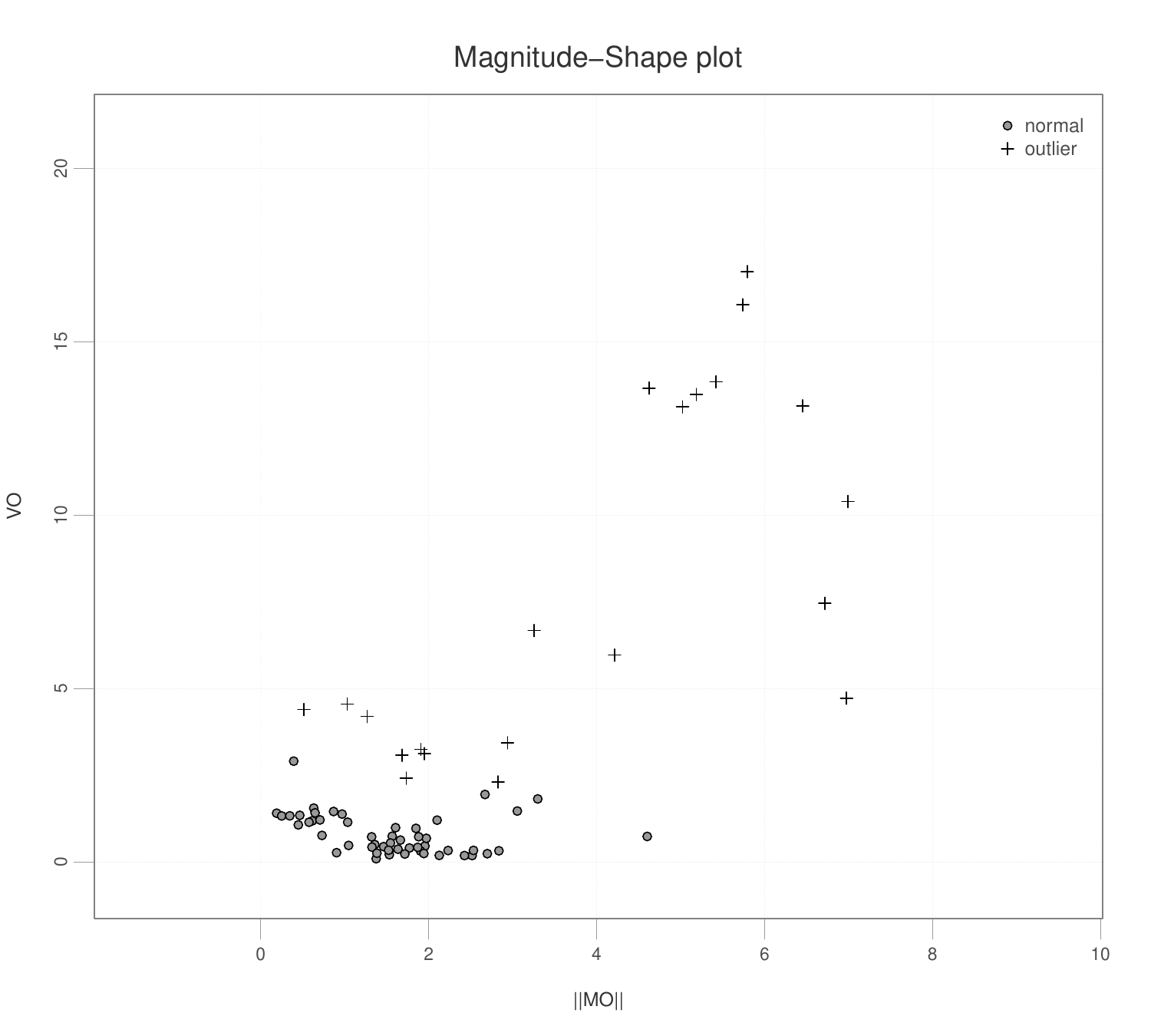}
\caption{\label{fig:aemet3} Plot of VO and $||\textbf{MO}||$ for the joint multivariate functional data  of temperature and log precipitation.}
\end{figure}

Figure \ref{fig:aemet3} shows the plot of the variation of outlyingness VO against the norm of
the mean of outlyingness $||\textbf{MO}||$ produced by \pkg{fdaoutlier} (by setting \code{plot = TRUE} in the call to \code{msplot()}).

Another to option to detect joint outliers is to use the directional outlyingness transformation $\mathcal{O}(\textbf{Y})(t)$ (together with a one-sided ERLD in the functional boxplot) in \cite{dai2020sequential}. This can be achieved using the \code{seq_transform()} function in \pkg{fdaoutlier}.
\begin{Schunk}
\begin{Sinput}
R> joint_seq <- seq_transform(dts = joint_dt, sequence = "O",
+                depth_method = "erld", erld_type = "one_sided_right")
R> joint_seq$outliers
\end{Sinput}
\begin{Soutput}
$O
 [1] 33 34 35 36 39 44 45 55 56 57 58 60 66
\end{Soutput}
\end{Schunk}

\begin{figure}[t!]
\centering
\includegraphics{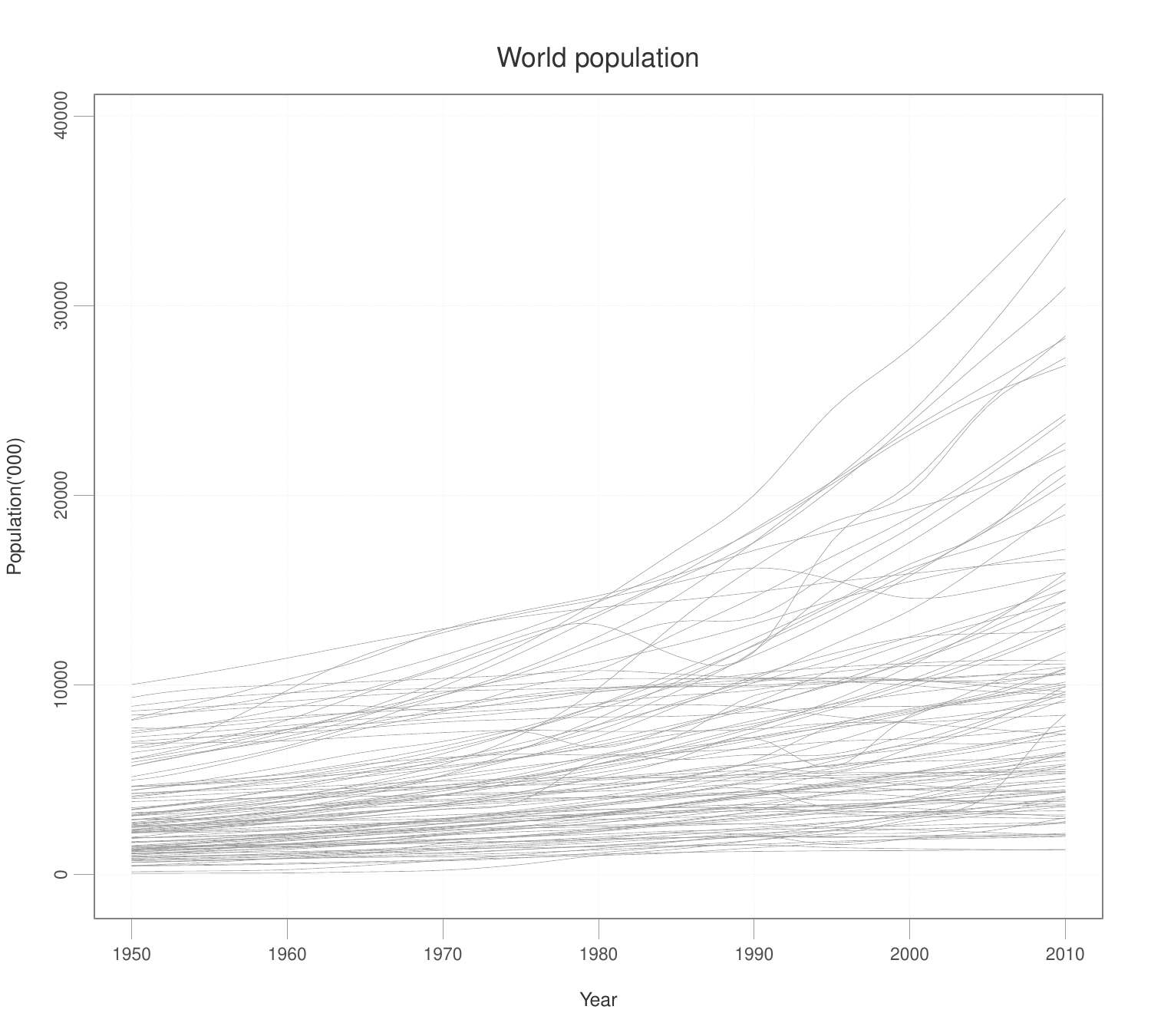}
\caption{\label{fig:worldpop1} World population in thousands of 105 countries from 1950-2010.}
\end{figure}

As a second example, we consider the world population data analysis carried out in \cite{dai2020sequential}. The data consists of the population of 105 countries as of July 1 in the years 1950-2010. These 105 countries are countries with a population between 1 million and fifteen million on July 1, 1980. The preprocessed data is available in \pkg{fdaoutlier} under the name \code{world_population}. A plot of the data is shown in Figure~\ref{fig:worldpop1}. Using the \code{seq_tranform()} function, we try to reproduce the results of the analysis carried out in \cite{dai2020sequential} which identified different types of outliers in the data using the the transformation $\mathcal{T}_2 \circ\ \mathcal{T}_1 \circ \mathcal{T}_0$ (and the $L^\infty$ depth in the functional boxplots).

\begin{figure}[t!]
\centering
\includegraphics{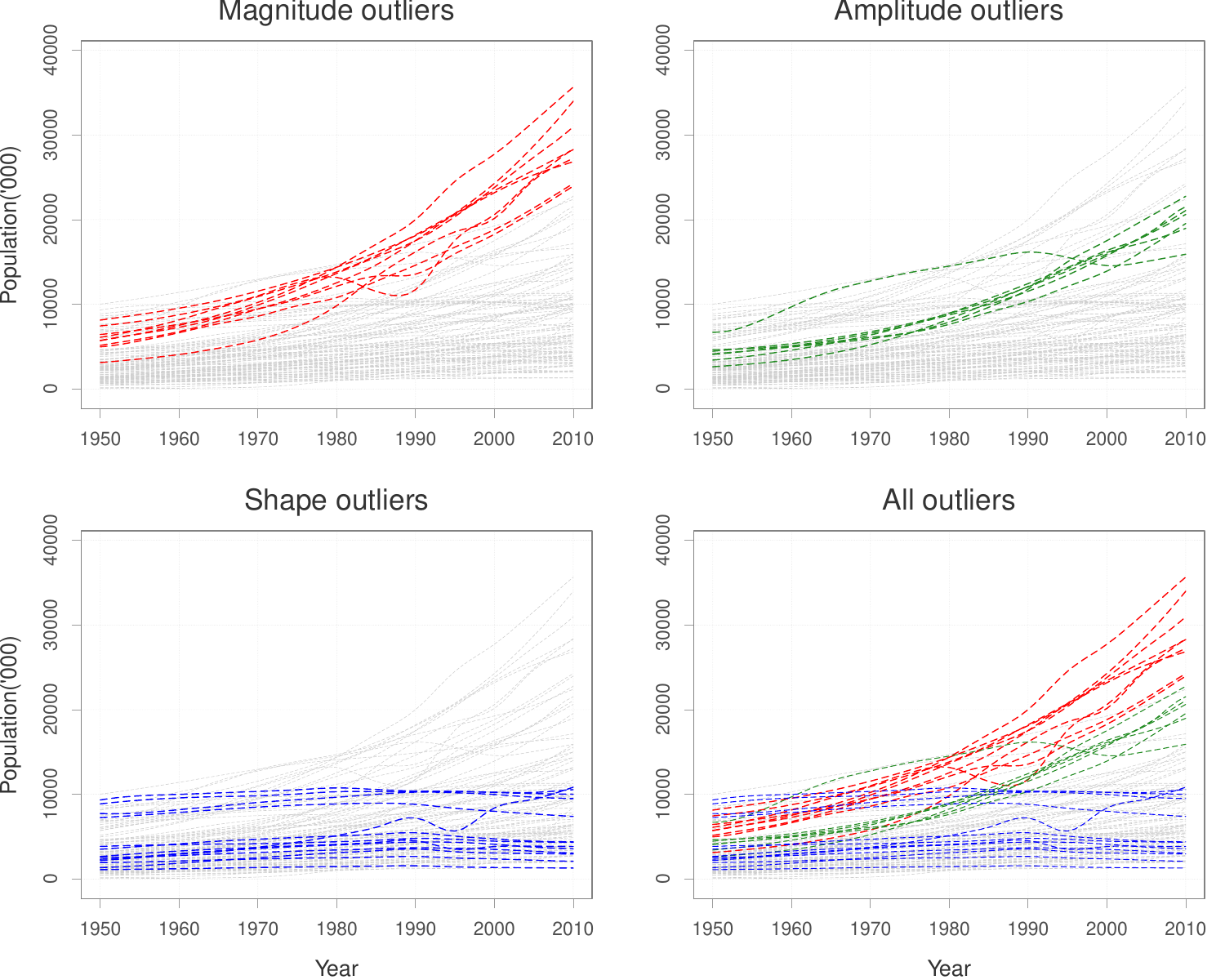}
\caption{\label{fig:worldpop2} Outliers detected in world population data using sequential transformation. Curves in red are magnitude outliers, in blue are shape outliers, in green are amplitude outliers and in grey are normal observations.}
\end{figure}

\begin{Schunk}
\begin{Sinput}
R> seq_pop <- seq_transform(dts = world_population,
+                           sequence = c("T0", "T1", "T2"),
+                           depth_method = "linfinity")
\end{Sinput}
\end{Schunk}
The magnitude outliers are then the $\mathcal{T}_0$~outliers which can be obtained with:
\begin{Schunk}
\begin{Sinput}
R> seq_pop$outliers$T0
\end{Sinput}
\begin{Soutput}
[1]  5  9 18 25 40 41 44 49 55
\end{Soutput}
\begin{Sinput}
R> (t0_outliers <- rownames(world_population)[seq_pop$outliers$T0])
\end{Sinput}
\begin{Soutput}
[1] "Mozambique"   "Uganda"       "Sudan"        "Ghana"       
[5] "Afghanistan"  "Nepal"        "Malaysia"     "Iraq"        
[9] "Saudi Arabia"
\end{Soutput}
\end{Schunk}
and the $\mathcal{T}_1$~outliers:
\begin{Schunk}
\begin{Sinput}
R> seq_pop$outliers$T1
\end{Sinput}
\begin{Soutput}
 [1]  3  5  9 12 13 18 24 25 36 40 41 44 49 55 57 59
\end{Soutput}
\begin{Sinput}
R> (t1_outliers <- rownames(world_population)[seq_pop$outliers$T1])
\end{Sinput}
\begin{Soutput}
 [1] "Madagascar"           "Mozambique"          
 [3] "Uganda"               "Angola"              
 [5] "Cameroon"             "Sudan"               
 [7] "Cote d'Ivoire"        "Ghana"               
 [9] "Kazakhstan"           "Afghanistan"         
[11] "Nepal"                "Malaysia"            
[13] "Iraq"                 "Saudi Arabia"        
[15] "Syrian Arab Republic" "Yemen"               
\end{Soutput}
\end{Schunk}
In \cite{dai2020sequential}, they considered the $\mathcal{T}_1$~outliers which are not $\mathcal{T}_0$~outliers as amplitude outliers for classification purposes. These can be obtained with:
\begin{Schunk}
\begin{Sinput}
R> amp_ind <- seq_pop$outliers$T1[!(seq_pop$outliers$T1
+                                   %in% seq_pop$outliers$T0)]
R> rownames(world_population)[amp_ind]
\end{Sinput}
\begin{Soutput}
[1] "Madagascar"           "Angola"              
[3] "Cameroon"             "Cote d'Ivoire"       
[5] "Kazakhstan"           "Syrian Arab Republic"
[7] "Yemen"               
\end{Soutput}
\end{Schunk}
Finally, the shape outliers are the  $\mathcal{T}_2$~outliers that are neither $\mathcal{T}_0$~outliers nor  $\mathcal{T}_1$~outliers.
\begin{Schunk}
\begin{Sinput}
R> shape_ind <- seq_pop$outliers$T2[!(seq_pop$outliers$T2
+                          %in% seq_pop$outliers$T0)]
R> shape_ind <- shape_ind[!(shape_ind %in% seq_pop$outliers$T1)]
R> rownames(world_population)[shape_ind]
\end{Sinput}
\begin{Soutput}
 [1] "Rwanda"                 "Armenia"               
 [3] "Georgia"                "Belarus"               
 [5] "Bulgaria"               "Czech Republic"        
 [7] "Hungary"                "Republic of Moldova"   
 [9] "Estonia"                "Latvia"                
[11] "Lithuania"              "Bosnia and Herzegovina"
[13] "Croatia"               
\end{Soutput}
\end{Schunk}
The $\mathcal{T}_0$ and $\mathcal{T}_1$~outliers are mostly countries in Africa and the Middle East while the shape outliers ($\mathcal{T}_2$~outliers) are mostly Eastern and Central European countries. The outliers detected are visualised in Figure~\ref{fig:worldpop2}, and these results are consistent with \citet[Table 5]{dai2020sequential}. We can also use \code{tvdmss()} and \code{muod()} on the world population data:
\begin{Schunk}
\begin{Sinput}
R> wptvdout <- tvdmss(dts = world_population)
R> wpmout <- muod(dts = world_population, cut_method = "boxplot")
\end{Sinput}
\end{Schunk}

TVD did not detect any magnitude outlier but MSS does find a couple of ``shape'' outliers all of which are in Africa and the Middle East, and these countries are those classified as $\mathcal{T}_0$ and $\mathcal{T}_1$~outliers by sequential transformations above:
\begin{Schunk}
\begin{Sinput}
R> wptvdout$magnitude_outliers
\end{Sinput}
\begin{Soutput}
NULL
\end{Soutput}
\begin{Sinput}
R> rownames(world_population)[wptvdout$shape_outliers]
\end{Sinput}
\begin{Soutput}
 [1] "Mozambique"           "Uganda"              
 [3] "Sudan"                "Cote d'Ivoire"       
 [5] "Ghana"                "Kazakhstan"          
 [7] "Afghanistan"          "Nepal"               
 [9] "Malaysia"             "Iraq"                
[11] "Saudi Arabia"         "Syrian Arab Republic"
[13] "United Arab Emirates" "Yemen"               
\end{Soutput}
\end{Schunk}
The countries flagged as magnitude outliers by \code{muod()} are very similar to those flagged by sequential transformations as $\mathcal{T}_0$~outliers:
\begin{Schunk}
\begin{Sinput}
R> rownames(world_population)[wpmout$outliers$magnitude]
\end{Sinput}
\begin{Soutput}
[1] "Sudan"         "Saudi Arabia"  "Uganda"        "Iraq"         
[5] "Cote d'Ivoire" "Malaysia"     
\end{Soutput}
\end{Schunk}
The amplitude outliers detected by \code{muod()} that are also not magnitude outliers are:

\begin{Schunk}
\begin{Sinput}
R> wpamp_ind <- wpmout$outliers$amplitude[!(wpmout$outliers$amplitude
+                                         %in% wpmout$outliers$magnitude)]
R> rownames(world_population)[wpamp_ind]
\end{Sinput}
\begin{Soutput}
[1] "Nepal"                "Ghana"               
[3] "Syrian Arab Republic" "Yemen"               
[5] "Afghanistan"          "Mozambique"          
[7] "Madagascar"          
\end{Soutput}
\end{Schunk}

Finally, the shape outliers flagged by \code{muod()} that are neither amplitude nor magnitude outliers are mostly the Eastern and Central European countries flagged as $\mathcal{T}_2$~outliers by sequential tranformations:
\begin{Schunk}
\begin{Sinput}
R> wpshape_ind <- wpmout$outliers$shape[!(wpmout$outliers$shape
+                                         %in% wpmout$outliers$magnitude)]
R> wpshape_ind <- wpshape_ind[!(wpshape_ind %in% wpmout$outliers$amplitude)]
R> rownames(world_population)[wpshape_ind]
\end{Sinput}
\begin{Soutput}
 [1] "Bulgaria"               "Hungary"               
 [3] "Latvia"                 "Georgia"               
 [5] "Estonia"                "Bosnia and Herzegovina"
 [7] "Lithuania"              "Republic of Moldova"   
 [9] "Croatia"                "Armenia"               
[11] "Belarus"                "United Arab Emirates"  
[13] "Kazakhstan"             "Czech Republic"        
\end{Soutput}
\end{Schunk}
In conclusion, the outliers detected and the classification of such outliers may vary across different outlier detection methods as shown by this world population data example. While the results of the outlier detection and classification for MUOD and sequential transformations are quite similar, those of TVD and MSS are quite different and in particular, countries identified as shape outliers by TVD and MSS are those flagged as magnitude and amplitude outliers by MUOD and sequential tranformations.


\section{Summary and future outlook} \label{sec:summary}

This article introduces the \pkg{fdaoutlier} package, which extends the available tools for outlier detection in functional data analysis in \proglang{R}. \pkg{fdaoutlier}'s focus so far has been the implementation of the latest state-of-the-art outlier detection methods in the FDA literature that are not yet implemented in \proglang{R} and that are especially useful for detecting shape outliers. These include the directional outlyingness and MS-Plot, the total variation depth and modified shape similarity, and sequential transformations. These implementations will be especially useful to FDA researchers for testing and comaprisons in the development of new outlier detection and exploratory methods. Likewise, \pkg{fdaoutlier} will be useful for practicioners in the exploratory analysis of their functional data.

We will continue adding more outlier detection methods for functional data to the \pkg{fdaoutlier} package, especially (future) outlier detection methods not yet implemented in \proglang{R}, and also those with implementations in \proglang{R} but that can be improved upon. We will also continue adding useful tools for development and testing of new outlier detection methods for functional data, e.g., additional simulation models and functions for comparing outlier detection methods, and these will be directed by the trends in the FDA literature. Our long-term goal for the development of \pkg{fdaoutlier} is for it to be a helpful package for practitioners and researchers alike for conducting robust analysis and outlier detection for functional data.


\section*{Computational details}

The results in this paper were obtained using
\proglang{R}~3.6.3 with the
\pkg{fdaoutlier}~0.2.0 package and \pkg{fda.usc}~2.0.2. \proglang{R} itself
and all packages used are available from the Comprehensive
\proglang{R} Archive Network (CRAN) at
\url{https://CRAN.R-project.org/}.

\section*{Acknowledgments}

This research is partially funded by the Regional Government of Madrid (CM) grant EdgeData-CM (P2018/TCS4499) cofunded by FSE \& FEDER, the Ministry of Science and Innovation grant PID2019-109805RB-I00 (ECID) cofunded by FEDER, and the Ministerio de Ciencia, Innovación y Universidades grant number PID2019-104901RB-I00.


\bibliography{refs}

\begin{thebibliography}{32}
\newcommand{\enquote}[1]{``#1''}
\providecommand{\natexlab}[1]{#1}
\providecommand{\url}[1]{\texttt{#1}}
\providecommand{\urlprefix}{URL }
\expandafter\ifx\csname urlstyle\endcsname\relax
  \providecommand{\doi}[1]{doi:\discretionary{}{}{}#1}\else
  \providecommand{\doi}{doi:\discretionary{}{}{}\begingroup
  \urlstyle{rm}\Url}\fi
\providecommand{\eprint}[2][]{\url{#2}}

\bibitem[{Arribas-Gil and Romo(2014)}]{arribas2014shape}
Arribas-Gil A, Romo J (2014).
\newblock \enquote{Shape Outlier Detection and Visualization for Functional
  Data: the Outliergram.}
\newblock \emph{Biostatistics}, \textbf{15}(4), 603--619.
\newblock ISSN 1465-4644.
\newblock \doi{10.1093/biostatistics/kxu006}.
\newblock \urlprefix\url{https://doi.org/10.1093/biostatistics/kxu006}.

\bibitem[{Azcorra \emph{et~al.}(2018)Azcorra, Chiroque, Cuevas,
  Fern{\'a}ndez~Anta, Laniado, Lillo, Romo, and
  Sguera}]{azcorra2018unsupervised}
Azcorra A, Chiroque LF, Cuevas R, Fern{\'a}ndez~Anta A, Laniado H, Lillo RE,
  Romo J, Sguera C (2018).
\newblock \enquote{Unsupervised Scalable Statistical Method for Identifying
  Influential Users in Online Social Networks.}
\newblock \emph{Scientific reports}, \textbf{8}(1), 1--7.
\newblock \doi{10.1038/s41598-018-24874-2}.
\newblock \urlprefix\url{https://doi.org/10.1038/s41598-018-24874-2}.

\bibitem[{Brys \emph{et~al.}(2005)Brys, Hubert, and
  Rousseeuw}]{brys2005robustification}
Brys G, Hubert M, Rousseeuw PJ (2005).
\newblock \enquote{A Robustification of Independent Component Analysis.}
\newblock \emph{Journal of Chemometrics}, \textbf{19}(5‐7), 364--375.
\newblock \doi{https://doi.org/10.1002/cem.940}.
\newblock
  \urlprefix\url{https://onlinelibrary.wiley.com/doi/abs/10.1002/cem.940}.

\bibitem[{Dai and Genton(2018)}]{dai2018multivariate}
Dai W, Genton MG (2018).
\newblock \enquote{Multivariate Functional Data Visualization and Outlier
  Detection.}
\newblock \emph{Journal of Computational and Graphical Statistics},
  \textbf{27}(4), 923--934.
\newblock \doi{10.1080/10618600.2018.1473781}.
\newblock \urlprefix\url{https://doi.org/10.1080/10618600.2018.1473781}.

\bibitem[{Dai and Genton(2019)}]{dai2019directional}
Dai W, Genton MG (2019).
\newblock \enquote{Directional Outlyingness for Multivariate Functional Data.}
\newblock \emph{Computational Statistics \& Data Analysis}, \textbf{131},
  50--65.
\newblock ISSN 0167-9473.
\newblock \doi{10.1016/j.csda.2018.03.017}.
\newblock
  \urlprefix\url{https://www.sciencedirect.com/science/article/pii/S016794731830077X}.

\bibitem[{Dai \emph{et~al.}(2020)Dai, Mrkvička, Sun, and
  Genton}]{dai2020sequential}
Dai W, Mrkvička T, Sun Y, Genton MG (2020).
\newblock \enquote{Functional Outlier Detection and Taxonomy by Sequential
  Transformations.}
\newblock \emph{Computational Statistics \& Data Analysis}, \textbf{149},
  106960.
\newblock ISSN 0167-9473.
\newblock \doi{10.1016/j.csda.2020.106960}.
\newblock
  \urlprefix\url{http://www.sciencedirect.com/science/article/pii/S0167947320300517}.

\bibitem[{Febrero \emph{et~al.}(2007)Febrero, Galeano, and
  Gonz{\'a}lez-Manteiga}]{febrero2007functional}
Febrero M, Galeano P, Gonz{\'a}lez-Manteiga W (2007).
\newblock \enquote{A Functional Analysis of NOx Levels: Location and Scale
  Estimation and Outlier Detection.}
\newblock \emph{Computational Statistics}, \textbf{22}(3), 411--427.
\newblock \doi{10.1007/s00180-007-0048-x}.
\newblock \urlprefix\url{https://doi.org/10.1007/s00180-007-0048-xx}.

\bibitem[{Febrero \emph{et~al.}(2008)Febrero, Galeano, and
  Gonz{\'a}lez-Manteiga}]{febrero2008}
Febrero M, Galeano P, Gonz{\'a}lez-Manteiga W (2008).
\newblock \enquote{Outlier Detection in Functional Data by Depth Measures, with
  Application to Identify Abnormal NOx Levels.}
\newblock \emph{Environmetrics}, \textbf{19}(4), 331--345.
\newblock \doi{10.1002/env.878}.
\newblock
  \urlprefix\url{https://onlinelibrary.wiley.com/doi/abs/10.1002/env.878}.

\bibitem[{Febrero-Bande and {Oviedo de la Fuente}(2012)}]{fda.usc}
Febrero-Bande M, {Oviedo de la Fuente} M (2012).
\newblock \enquote{Statistical Computing in Functional Data Analysis: The
  \proglang{R} Package \pkg{fda.usc}.}
\newblock \emph{Journal of Statistical Software, Articles}, \textbf{51}(4),
  1--28.
\newblock ISSN 1548-7660.
\newblock \doi{10.18637/jss.v051.i04}.
\newblock \urlprefix\url{https://www.jstatsoft.org/v051/i04}.

\bibitem[{Hardin and Rocke(2005)}]{hardinrocke3005}
Hardin J, Rocke DM (2005).
\newblock \enquote{The Distribution of Robust Distances.}
\newblock \emph{Journal of Computational and Graphical Statistics},
  \textbf{14}(4), 928--946.
\newblock \doi{10.1198/106186005X77685}.
\newblock \eprint{https://doi.org/10.1198/106186005X77685}.

\bibitem[{Huang and Sun(2019)}]{huang2019decomposition}
Huang H, Sun Y (2019).
\newblock \enquote{A Decomposition of Total Variation Depth for Understanding
  Functional Outliers.}
\newblock \emph{Technometrics}, \textbf{61}(4), 445--458.
\newblock \doi{10.1080/00401706.2019.1574241}.
\newblock \urlprefix\url{https://doi.org/10.1080/00401706.2019.1574241}.

\bibitem[{Hubert \emph{et~al.}(2015)Hubert, Rousseeuw, and
  Segaert}]{hubert2015multivariate}
Hubert M, Rousseeuw PJ, Segaert P (2015).
\newblock \enquote{Multivariate Functional Outlier Detection.}
\newblock \emph{Statistical Methods \& Applications}, \textbf{24}(2), 177--202.
\newblock \doi{10.1007/s10260-015-0297-8}.
\newblock \urlprefix\url{https://doi.org/10.1007/s10260-015-0297-8}.

\bibitem[{Hubert and Van~der Veeken(2008)}]{hubert2008outlier}
Hubert M, Van~der Veeken S (2008).
\newblock \enquote{Outlier Detection for Skewed Data.}
\newblock \emph{Journal of Chemometrics}, \textbf{22}(3‐4), 235--246.
\newblock \doi{10.1002/cem.1123}.
\newblock
  \urlprefix\url{https://onlinelibrary.wiley.com/doi/abs/10.1002/cem.1123}.

\bibitem[{Hyndman and {Shahid Ullah}(2007)}]{hyndmanullah2018}
Hyndman RJ, {Shahid Ullah} M (2007).
\newblock \enquote{Robust Forecasting of Mortality and Fertility Rates: A
  Functional Data Approach.}
\newblock \emph{Computational Statistics \& Data Analysis}, \textbf{51}(10),
  4942 -- 4956.
\newblock ISSN 0167-9473.
\newblock \doi{10.1016/j.csda.2006.07.028}.
\newblock
  \urlprefix\url{http://www.sciencedirect.com/science/article/pii/S0167947306002453}.

\bibitem[{Hyndman and Shang(2010)}]{hyndman2010bagplot}
Hyndman RJ, Shang HL (2010).
\newblock \enquote{Rainbow Plots, Bagplots, and Boxplots for Functional Data.}
\newblock \emph{Journal of Computational and Graphical Statistics},
  \textbf{19}(1), 29--45.
\newblock \doi{10.1198/jcgs.2009.08158}.
\newblock \urlprefix\url{https://doi.org/10.1198/jcgs.2009.08158}.

\bibitem[{Ieva and Paganoni(2020)}]{ieva2020component}
Ieva F, Paganoni AM (2020).
\newblock \enquote{Component-Wise Outlier Detection Methods for Robustifying
  Multivariate Functional Samples.}
\newblock \emph{Statistical Papers}, \textbf{61}(2), 595--614.
\newblock \doi{10.1007/s00362-017-0953-1}.
\newblock \urlprefix\url{https://doi.org/10.1007/s00362-017-0953-1}.

\bibitem[{Ieva \emph{et~al.}(2019)Ieva, Paganoni, Romo, and
  Tarabelloni}]{ieva2019roahd}
Ieva F, Paganoni AM, Romo J, Tarabelloni N (2019).
\newblock \enquote{\pkg{roahd} Package: Robust Analysis of High Dimensional
  Data.}
\newblock \emph{{The \proglang{R} Journal}}, \textbf{11}(2), 291--307.
\newblock \doi{10.32614/RJ-2019-032}.
\newblock \urlprefix\url{https://doi.org/10.32614/RJ-2019-032}.

\bibitem[{Long and Huang(2016)}]{long2015}
Long JP, Huang JZ (2016).
\newblock \enquote{A Study of Functional Depths.}
\newblock \urlprefix\url{https://arxiv.org/abs/1506.01332}.

\bibitem[{L{\'o}pez-Pintado and Romo(2009)}]{Romo_depths}
L{\'o}pez-Pintado S, Romo J (2009).
\newblock \enquote{On the Concept of Depth for Functional Data.}
\newblock \emph{Journal of the American Statistical Association},
  \textbf{104}(486), 718--734.
\newblock \doi{10.1198/jasa.2009.0108}.
\newblock \eprint{https://doi.org/10.1198/jasa.2009.0108}.

\bibitem[{Myllymäki \emph{et~al.}(2017)Myllymäki, Mrkvička, Grabarnik,
  Seijo, and Hahn}]{mari2017}
Myllymäki M, Mrkvička T, Grabarnik P, Seijo H, Hahn U (2017).
\newblock \enquote{Global Envelope Tests for Spatial Processes.}
\newblock \emph{Journal of the Royal Statistical Society B}, \textbf{79}(2),
  381--404.
\newblock \doi{10.1111/rssb.12172}.
\newblock
  \urlprefix\url{https://rss.onlinelibrary.wiley.com/doi/abs/10.1111/rssb.12172}.

\bibitem[{Nagy \emph{et~al.}(2017)Nagy, Gijbels, and Hlubinka}]{nagy07depth}
Nagy S, Gijbels I, Hlubinka D (2017).
\newblock \enquote{Depth-Based Recognition of Shape Outlying Functions.}
\newblock \emph{Journal of Computational and Graphical Statistics},
  \textbf{26}(4), 883--893.
\newblock \doi{10.1080/10618600.2017.1336445}.
\newblock \urlprefix\url{https://doi.org/10.1080/10618600.2017.1336445}.

\bibitem[{Pokotylo \emph{et~al.}(2019)Pokotylo, Mozharovskyi, and
  Dyckerhoff}]{pokotylo2019ddalpha}
Pokotylo O, Mozharovskyi P, Dyckerhoff R (2019).
\newblock \enquote{Depth and Depth-Based Classification with \proglang{R}
  Package \pkg{ddalpha}.}
\newblock \emph{Journal of Statistical Software, Articles}, \textbf{91}(5),
  1--46.
\newblock ISSN 1548-7660.
\newblock \doi{10.18637/jss.v091.i05}.
\newblock \urlprefix\url{https://www.jstatsoft.org/v091/i05}.

\bibitem[{Ramsay \emph{et~al.}(2020)Ramsay, Graves, and Hooker}]{fdapackage}
Ramsay JO, Graves S, Hooker G (2020).
\newblock \emph{\pkg{fda}: Functional Data Analysis}.
\newblock \proglang{R} package version 5.1.5.1,
  \urlprefix\url{https://CRAN.R-project.org/package=fda}.

\bibitem[{{\proglang{R} Core Team}(2020)}]{Rcore}
{\proglang{R} Core Team} (2020).
\newblock \emph{\proglang{R}: {A} Language and Environment for Statistical
  Computing}.
\newblock \proglang{R} Foundation for Statistical Computing, Vienna, Austria.
\newblock \urlprefix\url{https://www.R-project.org/}.

\bibitem[{Rousseeuw and Driessen(1999)}]{rousseeuw1999fast}
Rousseeuw PJ, Driessen KV (1999).
\newblock \enquote{A Fast Algorithm for the Minimum Covariance Determinant
  Estimator.}
\newblock \emph{Technometrics}, \textbf{41}(3), 212--223.
\newblock \doi{10.1080/00401706.1999.10485670}.
\newblock
  \urlprefix\url{https://www.tandfonline.com/doi/abs/10.1080/00401706.1999.10485670}.

\bibitem[{Rousseeuw \emph{et~al.}(2018)Rousseeuw, Raymaekers, and
  Hubert}]{rousseeuw2018measure}
Rousseeuw PJ, Raymaekers J, Hubert M (2018).
\newblock \enquote{A Measure of Directional Outlyingness With Applications to
  Image Data and Video.}
\newblock \emph{Journal of Computational and Graphical Statistics},
  \textbf{27}(2), 345--359.
\newblock \doi{10.1080/10618600.2017.1366912}.
\newblock \urlprefix\url{https://doi.org/10.1080/10618600.2017.1366912}.

\bibitem[{Segaert \emph{et~al.}(2020)Segaert, Hubert, Rousseeuw, and
  Raymaekers}]{mrfdepth}
Segaert P, Hubert M, Rousseeuw P, Raymaekers J (2020).
\newblock \emph{\pkg{mrfDepth}: Depth Measures in Multivariate, Regression and
  Functional Settings}.
\newblock \proglang{R} package version 1.0.13,
  \urlprefix\url{https://CRAN.R-project.org/package=mrfDepth}.

\bibitem[{Shang(2011)}]{hanlinshang2011rj}
Shang HL (2011).
\newblock \enquote{\pkg{rainbow}: An \proglang{R} Package for Visualizing
  Functional Time Series.}
\newblock \emph{{The \proglang{R} Journal}}, \textbf{3}(2), 54--59.
\newblock \doi{10.32614/RJ-2011-019}.
\newblock \urlprefix\url{https://doi.org/10.32614/RJ-2011-019}.

\bibitem[{Sun and Genton(2011)}]{sun2011functional}
Sun Y, Genton MG (2011).
\newblock \enquote{Functional Boxplots.}
\newblock \emph{Journal of Computational and Graphical Statistics},
  \textbf{20}(2), 316--334.
\newblock \doi{10.1198/jcgs.2011.09224}.
\newblock \urlprefix\url{https://doi.org/10.1198/jcgs.2011.09224}.

\bibitem[{Sun \emph{et~al.}(2012)Sun, Genton, and Nychka}]{sun2012exact}
Sun Y, Genton MG, Nychka DW (2012).
\newblock \enquote{Exact Fast Computation of Band Depth for Large Functional
  Datasets: How Quickly Can One Million Curves be Ranked?}
\newblock \emph{Stat}, \textbf{1}(1), 68--74.
\newblock \doi{10.1002/sta4.8}.
\newblock
  \urlprefix\url{https://onlinelibrary.wiley.com/doi/abs/10.1002/sta4.8}.

\bibitem[{Vinue and Epifanio(2020{\natexlab{a}})}]{ada}
Vinue G, Epifanio I (2020{\natexlab{a}}).
\newblock \emph{\pkg{adamethods}: Archetypoid Algorithms and Anomaly
  Detection}.
\newblock \proglang{R} package version 1.2.1,
  \urlprefix\url{https://CRAN.R-project.org/package=adamethods}.

\bibitem[{Vinue and Epifanio(2020{\natexlab{b}})}]{vinue2020robust}
Vinue G, Epifanio I (2020{\natexlab{b}}).
\newblock \enquote{Robust Archetypoids for Anomaly Detection in Big Functional
  Data.}
\newblock \emph{Advances in Data Analysis and Classification}, pp. 1--26.
\newblock \doi{10.1007/s11634-020-00412-9}.
\newblock \urlprefix\url{https://doi.org/10.1007/s11634-020-00412-9}.

\end{thebibliography}


\newpage

\begin{appendix}

\end{appendix}


\end{document}